\documentclass{llncs}
\usepackage{llncsdoc}
\usepackage{graphicx}
\usepackage{amsmath}
\usepackage{cite}
\begin{document}
	
	\title{Business Matter Experts do Matter: A Model-Driven Approach for Domain Specific Process Design and Monitoring}

	\author{Adrian Mos, Mario Cortes-Cornax}
	\institute{Xerox Research Center, 6 Chemin de Maupertuis, Meylan, France \\ \email{adrian.mos@xrce.xerox.com, mario.cortes@xrce.xerox.com}}

	\maketitle
	
	\begin{abstract}
		
		Business process design and monitoring are essential elements of Business Process Management (BPM), often relying on Service Oriented Architectures (SOA). However the current BPM approaches and standards have not sufficiently reduced the Business-IT gap. Today's solutions are mostly domain-independent and platform-dependent, which limits the ability of business matter experts to express business intent and enact process change. In contrast, the approach presented in this paper focuses on BPM and SOA environments in a domain-dependent and platform-independent way. We propose to add a domain specific-layer on top of current solutions so business stakeholders can design and understand their processes in a more intuitive way. We rely on previously proposed technical solutions and integrate them in an end-to-end methodology (from design to monitoring and back). The appropriateness and the feasibility of the approach is justified through a use case and a complete prototype implementation.
		
	\end{abstract}
	
	
	\section{Introduction}\label{sec:introduction}
	
	Business process design connected to execution and monitoring are critical for successful Business Process Management (BPM)~\cite{rosemann2015six}. Today, the Business Process Model and Notation~\cite{omg2011notation} (BPMN 2.0) has become the de-facto standard for business process modelling. With the aim at filling the Business-IT gap, significant effort has been put into bringing BPMN executable and closer to Service Oriented Architectures (SOA). A BPM Suite (BPMS) manages the process execution directing SOA calls to the appropriate services and generally provides monitoring infrastructure. While these components help alleviate agility problems that business stakeholders encounter, there are important limitations to the current approaches. We observed that most of the existing solutions are domain-independent and platform-dependent, which limit the power of business matter experts at the design and monitoring stages.
	
	Concerning \textbf{process design limitations}, the BPMN standard lacks guidance to reach executable processes from high-level process models. Silver~\cite{silver2009bpmn} highlights this problem, and proposes a level-based top-down approach to design business processes (\emph{Descriptive level}, \emph{Analytical level} and \emph{Execution level}). However, the generality of the most common BPMN 2.0 graphical elements, in particular the Task element, reduces semantic expressiveness~\cite{moody2009physics}. Business analysts require dedicated means (e.g., specific type of task with implicit domain knowledge) to effectively model their business domain (ex. logistics, healthcare, transportation, etc.)~\cite{pinggera2010structuring}. Domain Specific Languages (DSLs) are an effective means to deal with these problems, providing improvements in expressiveness and ease of use~\cite{mernik2005and}. More specifically, Domain Specific Process Modelling Languages (DSPMLs)~\cite{jablonski2009evolution} permits business stakeholders to design their processes in a much more intuitive way than BPMN.
	
	Regarding \textbf{monitoring limitations}, BPMS solutions collect and present data at the level of the process description, which is generic. This fact results in monitoring information that is collected in a generic way with respect to the business domain (ex., ``activity", ``gateway" or ``event") with no correlation with the business concepts (ex.``order handling" or ``shipping") apart from the simple matching ``label - BPMN element". This causes a number of problems: (1) it is hard to make use of the monitoring data in order to present meaningful metrics for business users, without significant configuration efforts for each BP; (2) it is difficult to correlate the business concepts to the execution of services in the SOA layer; (3) it is difficult to set wide-ranging SLAs that affect all BPs in the organization equally. For instance, it may be necessary to specify that all the ``shipping" operations, regardless the BP in which they occur, must execute in less than 2 days.
	
	In this paper, we present an approach that focuses on BPM and SOA environments in a domain-dependent and platform-independent way. Previous technical solutions~\cite{mos2013improving, mos2015domain, mos2013platform} are combined to present a methodological, model-driven approach that integrates domain specific modelling with domain specific monitoring in an end-to-end solution. The appropriateness and the feasibility of our approach is shown through a use case and a complete prototype implementation. The rest of the paper is structured as follows. Section~\ref{sec:overviewApproach} describes a general overview of our method, based on a running example. Section~\ref{sec:proposition} details the steps of the approach. Section~\ref{sec:prototypeAndValidation} focuses on the prototype implementation. Section~\ref{sec:relatedWork} presents related work and finally, Section~\ref{sec:conclusion} concludes and discusses future work.

	
	\section{Overview of the Approach}\label{sec:overviewApproach}

\begin{figure}
					\centering
					\vspace{-1em}
					\includegraphics[width=1\linewidth]{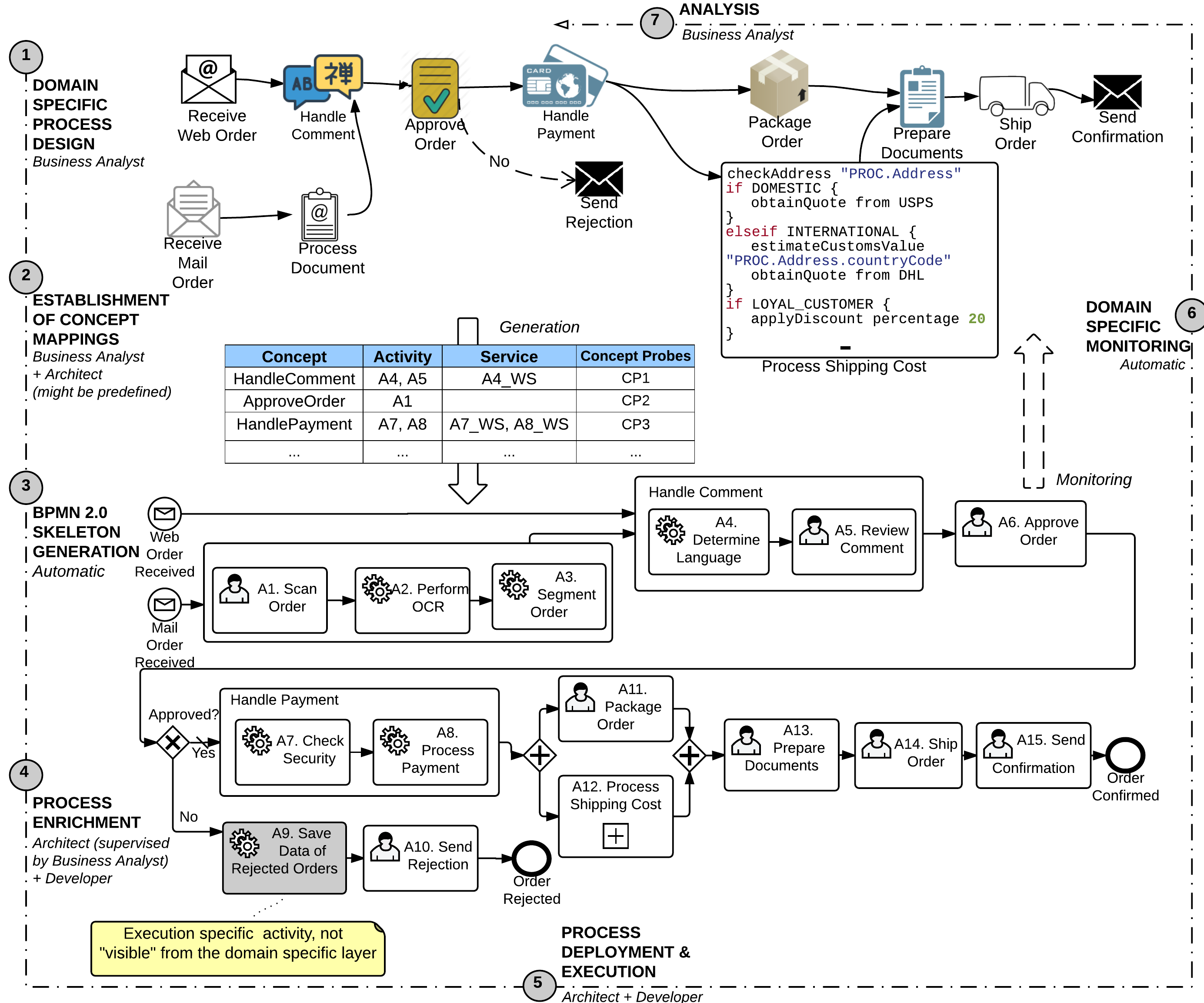}
					\vspace{-1em}
					\caption[Approach Overview]{Approach Overview with Main Steps}
					\vspace{-1em}
					\label{fig:approachOverview}
				\end{figure}

		
	Figure~\ref{fig:approachOverview} gives an overview of the approach from a modeller (business analyst and architect) point of view. Each number corresponds to one key step that will be further described in the following sections. The figure contains a simplified order handling process. The orders are received either by a submission web form or by standard mail. In the latter case, some document pre-processing is necessary in order to handle the order (i.e., scanning, Optical Character Recognition (OCR), and segmentation to extract the different sections). The order's comments, which could be in different languages, need to be handled before the approval. Afterwards, some classical processing steps such as payment, packaging, preparation of the documents (i.e., tracking number, bill), as well as the actual shipping and the confirmation are defined. In dotted lines, the business stakeholder indicates the exceptional paths. Each symbol represents a \emph{Domain Concept}, which makes reference to an enterprise well defined know-how element.
	
	The first step corresponds to the \textbf{domain specific design}, using a DSPML (\emph{Step~1} in Fig.~\ref{fig:approachOverview}). The domain specific language must have been previously designed based on the generic domain meta-model that we propose. Potentially, several DSPMLs can be combined, as the example shows. For instance, in order to define the process of calculating the shipping cost, a textual description may be more appropriate. In the graphical part, we advocate taking into account the principles of notations defined by Moody~\cite{moody2009physics}. More details about a particular language are out of the scope of this paper. However, rich language definitions are possible for various domains, as we show also in~\cite{mos2013improving}. 
	
	The analyst can then \textbf{establish the concept mappings} (\emph{Step~2} in Fig.~\ref{fig:approachOverview}). While business concepts are already connected by default to the abstract services from the enterprise repository, the links can still still be modified at this stage. This is essential in grounding the domain knowledge in technical realities. For instance, ``Handle Payment" corresponds to two technical services. It will imply the creation of the corresponding service tasks in BPMN. The mapping between domain concepts with the process activities relies on a pivot meta-model and unique ID (UID) attributes. The so-called Common meta-model (CommonMM) is a central, simplified representation of the main generic process concepts common to business process descriptions, such as activity, flow and gateway. It is significantly simpler than fully-fledged BPMN because its objective is simply to extract the essence of the structure of various business processes. Our hypothesis is that a descriptive level~\cite{silver2009bpmn} (reduced amount of symbols but semantically enriched) is enough to define high-level domain-specific process models.
	
	The \textbf{BPMN 2.0 skeleton is generated} relying on the aforementioned CommonMM (\emph{Step~3} in Fig.~\ref{fig:approachOverview}) and the concept mappings (see table with concepts mapped to activities). Note that the concept-to-activity mappings are generated or validated at this stage. Also note that the transformations are transparent to the business analyst. At most, the latter will have to agree with the business architect on the correspondence between the domain concepts and the to-be activities supported by generation templates. Once transformed into an instance of the CommonMM, the processes can be converted to the process modelling language of choice.
	
	\textbf{Generated BPMN models are typically enriched and refined} (\emph{Step~4} in Fig.~\ref{fig:approachOverview}). Extra activities, a complete data model or a resource model may be necessary in order to enable executability.
			
	\textbf{Deployment and execution} follow (\emph{Step~5} in Fig.~\ref{fig:approachOverview}). The only constraint that we impose here is the preservation of the concept mappings (i.e., not manually deleting the generated UIDs). Extra activities that may be added in the BPMN are considered as technical additions and of reduced interest from a business point of view (ex. activity A9). These activities will not be represented at the DSPML level when showing information coming from the domain specific monitoring infrastructure. The deployment phase is necessary to install the process artefacts and bind the generated abstract services with actual services, which will be running for instance in an enterprise service bus.
	
	\textbf{Monitoring} (\emph{Step~6} in Fig.~\ref{fig:approachOverview}) aims at aggregating and displaying data in the domain specific environment relying on information from the concept mappings. Our proposition aims to address the aforementioned shortcomings of today's monitoring capabilities for BPMS/SOA applications. A layer of abstraction is added on top of the existing capabilities rather than replacing them. The platform-independence ensures compatibility with a wide range of existing systems and platforms.
	
	Finally, in an \textbf{Analysis} stage (\emph{Step~7} in Fig.~\ref{fig:approachOverview}) the monitored data is studied to iteratively improve the process. Iteration may also imply the enrichment of the enterprise know-how, which is capitalised through the domain concepts.
	
	To summarise, the interest of the contribution is twofold: 1) the approach takes into account in a very specific way the business stakeholders, enabling domain specific modelling and monitoring; and 2) the entire cycle is integrated in a continuous improvement approach, supported by tools through model-driven transformations.
	
	
	\section{A Model-driven Approach for Domain Specific Process Design and Monitoring}\label{sec:proposition}
	
	This section details the main ideas of our model-driven approach for domain specific process design and monitoring, which considers business stakeholders as first class-citizens for BPM. The section mainly focuses on \emph{domain-specific design}, \emph{establishment of concept mappings} and \emph{domain-specific monitoring} which are the most relevant part of the work. The \emph{BPMN generation}, the \emph{process enrichment}, the \emph{deployment and execution} and the \emph{analysis}, while implemented and integrated, are not described in much detail, as they are relatively common BPM activities.
	
	
	\subsection{Domain-Specific Design through Domain Concepts}\label{sec:domainSpecificModelling}
	
	The interest and the limits of DSPMLs have already been presented in previous work~\cite{mos2015domain}. Naturally, our goal is not to propose a particular DSPML as their aim is to be adapted to particular business needs. Instead, we propose a generic domain description meta-model (MM), which provides a structural view of the domain. We then illustrate it with examples corresponding to the use-case described in the previous section. 
	
\begin{figure}
		\centering
		\vspace{-1em}
		\includegraphics[width=1\linewidth]{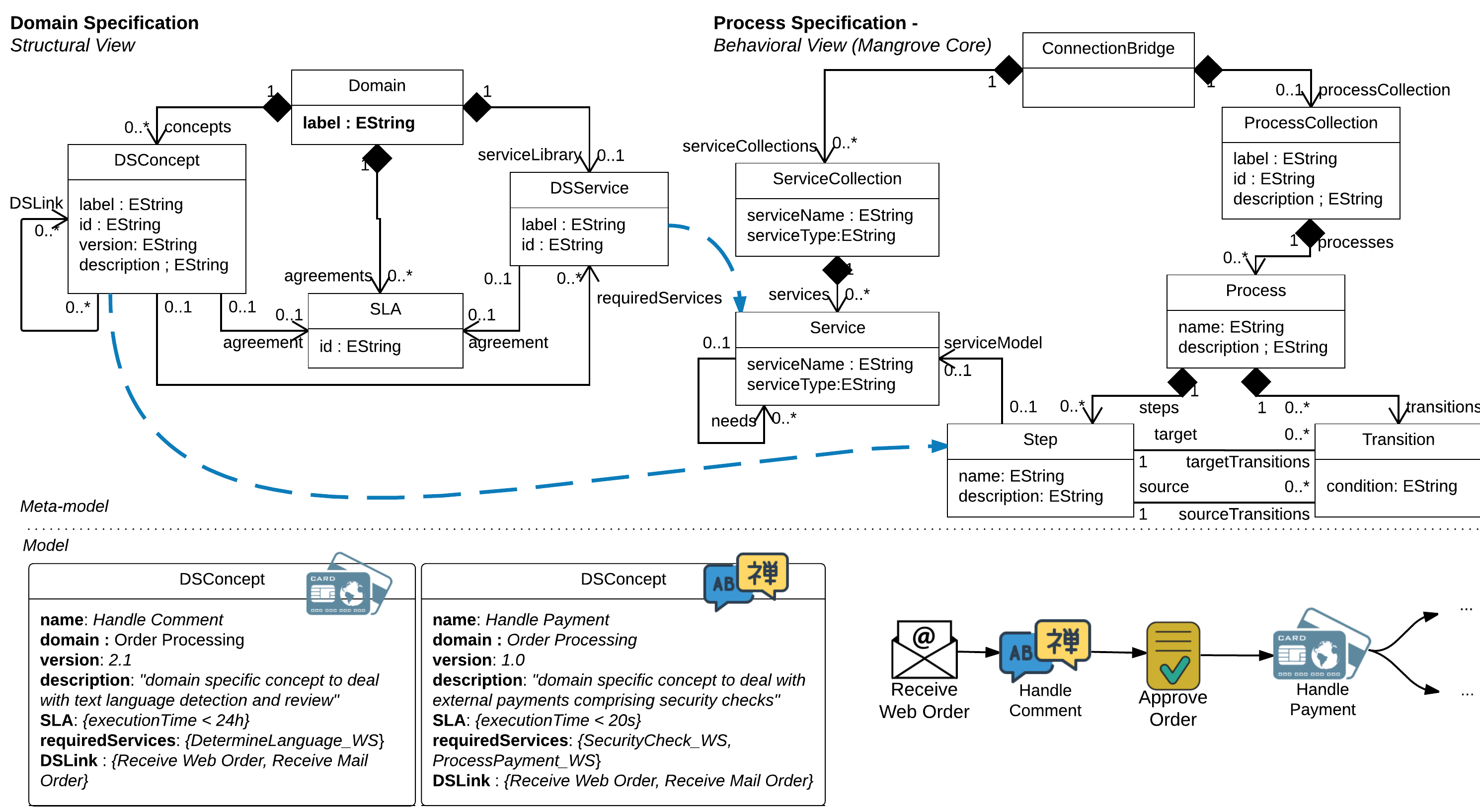}
		\vspace{-1em}
		\caption[DSConcepts]{Domain Specific Concepts Design Relies on a Generic Domain Meta-model}
		\vspace{-1em}
		\label{fig:dsConcepts}
	\end{figure}

	The upper part of Fig.~\ref{fig:dsConcepts} provides in a simplified way, the meta-models used to define the key points of the business domain in a generic way. They represent business domain information for an enterprise, with regard to the specification of concepts that are going to be reused in the business processes. The domain meta-model is useful for several proposes: 1) to store the domain information in a central repository on a collaboration and distribution server. This allows common access to the defined concepts to all the business users; 2) to generate a domain editor (textual) that can be used stand-alone or embedded in a graphical editor as part of a diagram designer; 3) to make the connection with the pivot meta-model specifying how process steps are going to be represented. This point is important when in a diagram, the user specifies that a process step is going to perform a business function corresponding to a business concept; 4) to inform and update SLA for business concepts. An enterprise-wide SLA management ensures that all activities and all processes that refer to a particular business concept would be marked with appropriate SLA constraints. This can bring important advantages when changes to company policies have sweeping implications on many SLAs, as they can be automatically propagated to all the relevant activities and processes. 
	
	The meta-model in Fig.~\ref{fig:dsConcepts} defines a \emph{Domain}, which contains a set of domain specific concepts (\emph{DSConcept}). A \emph{Domain} also contains \emph{SLA} elements, describing the agreement details. A \emph{DSConcept} relates to \emph{DSService} elements describing the actual SOA services required in the domain. Note that services can be abstract entities bound later in the deployment phase~\cite{jacquin2015deployment}.
	
	\textbf{Illustration based on the use case}. A Domain Concept supports the representation of business domain knowledge in an enterprise. Fig.~\ref{fig:dsConcepts} illustrates in the bottom part how knowledge common to the enterprise is stored in two example domain concepts (in contrast to a pure BPMN approach where such information would be implicit in the minds of the designers). These concepts would typically be stored in shared repositories. The information comprises for instance a name (verb+object) a version number and the SLA. Links between domains concepts are defined in order to define dependencies. For example, the concepts: ``Handle Payment" and ``Handle Comment" are related to the ``Receive Mail Order" and the ``Receive Web Order" business activities. This means that a common DataObject will be shared between the BPMN activities that are generated. Note that the data-model generation is currently not supported by our solution, although it is being investigated. However, these links provide necessary hints to the architects and analysts that enrich the generated BPMN skeleton.
	
	
	\subsection{Establishment of Concept Mappings}\label{sec:conceptMappings}
	
	In their simplest form, concept mappings are connections between business concepts and the SOA services that are used by them. This relation, could be defined by means of process activities. A simple example of the concept mappings is presented in Fig.~\ref{fig:dsConcepts}. Concept mappings are defined as following:
	
	\begin{itemize}
		\item Set of services $S = \{ s_{1}, s_{2},...,s_{q}\}$
		\item Set of processes $P = \{ p_{1}, p_{2},...,p_{m}\}$
		\item For each process $p_{k}$, a set of activities $Ak = \{ a_{k1}, a_{k2},...,a_{kt(k)}\}$ where the number of the activities in the set \($t(k)$\) depends on the complexity of $p_{k}$
		\item The set of all activities in all processes $A = A_{1} \bigcup A_{2}\bigcup ... \bigcup A_{| P |}$
	\end{itemize}
	
	The goal of concept mappings is to determine the following sets:
	
	\begin{itemize}
		\item Set of concepts $C = \{ c_{1}, c_{2},...,c_{n}\}$
		\item $Concept Mappings (CM) = \{ c_{j}, s_{j} : \forall c_{j} \in C; S_{j} \subseteq S \}$ which contains for each concept its list of services, e.g., $HandlePayment, (s_{1}, s_{2})$.
		\item $Activity Mappings (AM_{k})= \{ a_{k}, c_{j} : \forall a_{ki} \in A_{k}; c_{j} \in C \}$ which contains for process $p_{k}$ its activities and the concepts they map to.
		\item $AM = AM_{1} \bigcup AM_{2} \bigcup ... \bigcup AM_{| P |}$ which contains for each activity all processes the concept it maps to.
	\end{itemize}
	
	Obtaining the sets $C$ and $CM$ requires that the business concepts used in the processes be clearly identified together with their required SOA services. Concepts are defined by business experts, connected to abstract services initially and eventually bound to real SOA services in the deployment stage as discussed in Section~\ref{sec:deploymentAndExecution}. The modelling environment needs to propose to the business expert a set of relevant SOA services. Other approaches, more or less automatic, for concept mapping could be applied~\cite{mos2015domain}. Once the concepts have been identified, it is necessary to obtain the AM set by mapping the BP's activities to the concepts (typically done automatically at the BPMN generation phase).
	
	\textbf{Illustration based on the use case.} A concept can have an immediate correspondence with a process activity (ex. ``Approve Order") or several activities (ex. ``Handle Comment", which refers to \emph{Determine Language} and \emph{Review Comment} in Fig.~\ref{fig:approachOverview}). A domain concept which is described with a textual DSL as for example the ``Process Shipping Cost", corresponds to a sub-process that will generate several BPMN activities. The sub-process itself contains a number of domain concepts that correspond to the knowledge about price management. These correspondences will vary depending on the enterprise domain concepts. In fact, the freedom to define such mappings brings an important level of flexibility in how business knowledge gets transferred into processes that are governed in a uniform way at the business level.
	
	\subsection{BPMN Skeleton Generation}\label{sec:bpmnGeneration}
	
	The BPMN generation relies on a Common Meta-model (CommonMM), which is a simplified representation of the main generic process concepts. The reason why the DSPML-based processes are not directly transformed into the generic language is to introduce flexibility in the approach. As the generic language (usually BPMN) evolves, only the transformation between the CommonMM and the target MM needs to be updated. It could be argued that the use of a simplified version of the BPMN meta-model, where only the descriptive objects are included, could facilitate the transformation process. However, if we aim to strictly follow the BPMN~2.0 meta-model, a complex class hierarchy should be respected. This particularity may not be shared with other process languages and would complicate transformations (ex. a \emph{Task} element subsequently inherits from \emph{Activity}, \emph{Flow Node}, \emph{Flow Element}, and \emph{Base Element}). For our prototype, we use Mangrove Core\footnote{http://www.eclipse.org/proposals/mangrove/} as our CommonMM (a simplified version is depicted in the upper right side in Fig.\ref{fig:dsConcepts}). Mangrove Core is a meta-model that unifies business processes and SOA elements. It provides behavioural support to the domain definition in order to define the necessary steps in a process. This framework, does not aim to manage a large collection of processes, such as APROMORE~\cite{delarosa2011apromore}. Instead, it focuses on preserving the sync between the common elements of business processes and architectural constructs from the various related diagrams.
	
			\begin{figure}
				\centering
				\vspace{-1.2em}
				\includegraphics[width=0.79\linewidth]{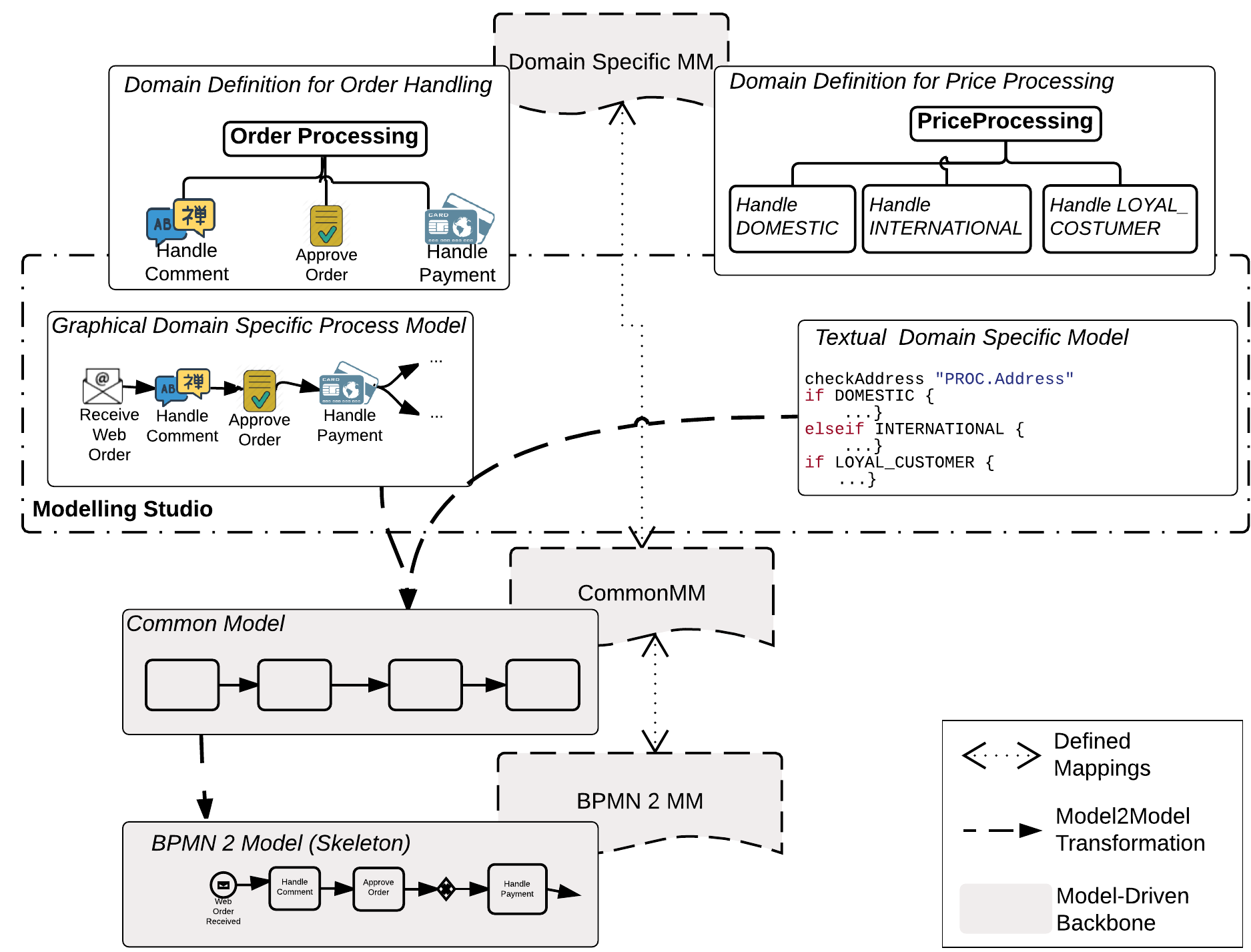}
				\vspace{-1em}
				\caption[Model Transformaitons]{Model Transformation from Domain Specific Models to BPMN 2.0 Model}
				\vspace{-1em}
				\label{fig:modelTransformations}
			\end{figure}
	
	In our approach, several target languages can be supported incrementally over time. When a new target is added, new transformations need of course to be added between the CommonMM and the new MM. We do not go into the details about the transformation process as the paper focuses on this general methodology. More details about the two-way synchronization between domain specific models and BPMN are presented in~\cite{mos2013improving}.
		
Figure~\ref{fig:modelTransformations} depicts how the BPMN generation is performed through model transformations. In our running example, two domain specific models are merged in a unified BPMN model. This shows the capacity of adaptation of the approach to different business expressibility needs. Both domain specific models leverage the DomainMM, which is mapped (i.e., MM concepts are linked) to the CommonMM. The latter is mapped to the target meta-model (in this case BPMN). The depicted meta-models provide a model driven backbone, where different domain specific models can be plugged in. The modelling studio (see Fig.~\ref{fig:modelTransformations}) is the tool that permits the business analysts to build specific process models connecting the predefined domain concepts using various process representations, based on their specific business domain.
	
	\subsection{Process Enrichment}\label{sec:enrichment}
	This stage relates to the need of the generated BPMN skeleton to be enriched if execution is targeted. New activities (ex. A9 in our running example), specific gateways and events, as well as several details may need to be added to the process model. We do not go into much details here as our approach does not impose any significant restrictions to this stage. The only constraint that the approach brings is to preserve the generated activities (tasks or sub-processes), so the link between the domain concepts and the process activities be maintained. Indeed, we did not force a perfect vertical alignment that could be very costly and unrealistic as described in~\cite{weidlich2009vertical}. The double synchronization mechanism explained in previous work~\cite{mos2013improving} permits to make (and propagate) changes in the domain model as well as in the generated BPMN model. The tracking of generated elements can be based on several identity-preserving mechanisms, of which a simple example is the usage of unique IDs injected in hidden properties of BPMN elements. This mechanism enables the possibility to make changes in the domain model as well as in the generated model.
	
	\subsection{Process Deployment and Execution}\label{sec:deploymentAndExecution}
	
	Concerning deployment, when defining business processes, individual business process activities can be connected to the service-execution capabilities of the enterprise, thus allowing any business process to be easily translated into an executable workflow on the platform of choice. This capability is enabled in our approach by providing mappings for each domain concept in order to specify how it should be grounded in the SOA. These mappings are done with idealized or abstract services in a two-step mechanism, in order to ensure better portability (and reusability) across the enterprise, as well as encourage proper adoption of good SOA-practices in future evolutions of the enterprise SOA. These abstract services (AS) would then be further connected to the real services in the repositories. The creation of these mappings would typically be performed by IT experts that have a good understanding of the domain and who envisage an ideal connection to a SOA. These abstract, idealised services, would not necessarily correspond on a 1 to 1 basis with business concepts as we show in the example. That is because there are important differences in concerns when defining business elements and when defining the service infrastructure, due to varying needs for reusability, performance and evolution of these two layers. In our approach, this two-step binding mechanism explained in \cite{mos2013platform} is applied to link domain concepts to any number of AS first and then each AS to real SOA services. 
	
	In order to execute the process, BPMS usually need at least a data-model defining the artefacts that flow in the process, a resource-model establishing the links between the roles defined in the process and actual users and the implementation of gateway conditions (usually based on data). These artefacts can be partly generated by the presented approach, but they may need to be enhanced by technical architects.
	
	\subsection{Domain-Specific Monitoring}\label{sec:monitoring}
	
	The main elements involved in domain specific monitoring are the Concept Probes (CPs) and the Business Process Probes (BPPs). There is a one-to-one correspondence between CPs and domain concepts. CPs collect an arbitrary number of metrics, such as execution time or execution status from the activities that are mapped to a domain concept. Once the CPs are created, they need to be bound to the monitoring capabilities of the existing infrastructure, effectively acting as an extra monitoring layer on top of the actual BPMS and SOA platforms. BPPs aggregate data from the BPMS and the various CPs. In order to enable them, they have to be linked to the domain concepts at design time (which is performed automatically). When all the required mappings are available, the probes are created, instantiated and deployed automatically respecting a predefined template. More technical details about concept probes can be found in~\cite{mos2015domain}. Here, we summarise their main functionality. Both CPs and BPPs are divided in three main components with particular concerns: the \emph{Raw Data Collection Component}, mainly collects data from the activities corresponding to each concept and the related technical services. The \emph{Analysis Component} is in charge of the aggregation of a raw data into composite metrics. These composite metrics are data structures that present the aggregate monitoring information combining the individual metric data for BPMS, SOA and other collection points such as Network Monitoring, App Server Monitoring and Operating System Monitoring. Finally, the \emph{Alerts Reporting Component} allows the registration of SLA requests through a configurable alert port. It uses the analysis component to constantly compare the aggregated metric values with the required thresholds.
	
	The approach provides the business stakeholders with means to govern their processes at a high level, with impact to the entire collection of business processes in a domain, if required. Relying on domain concepts, they are able to consistently manage the execution parameters of a large collection of process descriptions and their instances. For example, if the Shipping concept is already defined, it is automatically reused in any process description detailing shipping operations, carrying over the reuse of the generation and the monitoring infrastructure. The definition of corporate-level SLA is easily implemented and maintained. Relying on the generative approach, changes are spread through the different layers. In the long run, the monitoring mechanisms enable better decision making, based on domain specific information, by putting the appropriate level of information in the tools used by the business-matter experts. Section~\ref{sec:prototypeAndValidation} discusses the prototype implementation and provides more details on the actual set up of the monitoring probes.
	
		
	\subsection{Analysis}\label{sec:analysis}
	
	In order to close the iterative lifecycle loop depicted in Fig.~\ref{fig:approachOverview}, an analysis step  is necessary (\emph{Step~7} in the figure), where the analysts study the monitored data in order to improve the process. The novelty in our approach is that the new know-how acquired in the enactment of the process may imply the update or creation of domain concepts. One of the biggest advantages of the approach is that if an updated concept is being used in a collection of processes, the changes will more easily propagated through the complete stack.

	
	\section{Prototype Implementation and First Validation Steps}\label{sec:prototypeAndValidation}
	
\begin{figure}
				\centering
				\vspace{-1em}
				\includegraphics[width=0.7\linewidth]{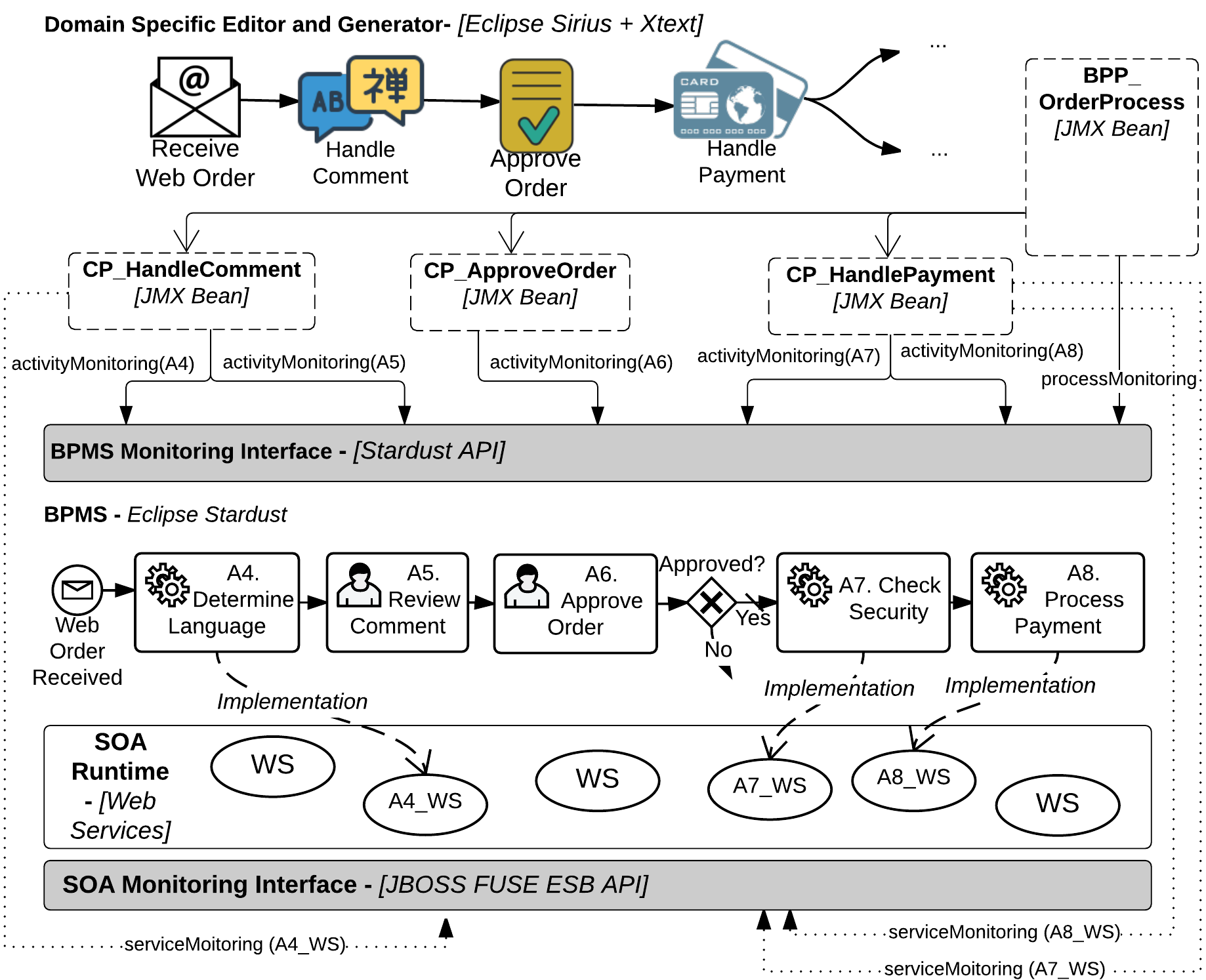}
				\vspace{-1em}
				\caption[PrototypeArchitecture]{Prototype Architecture}
				\vspace{-1.3em}
				\label{fig:prototypeArchitecture}
			\end{figure}

	Fig.~\ref{fig:prototypeArchitecture} depicts the architecture of the prototype illustrated for our use case. The picture shows the domain specific layer as an additional layer to the BPM and SOA stack. A domain specific editor would be the entry-point for a business stakeholder, providing domain specific process design (based on domain concepts), BPMN generation (which is transparent to business stakeholders) and display of monitoring result (outcome of the concept and process probes). We present some key points of the prototype implementation supporting the process life-cycle. This prototype is mostly based on Eclipse technologies, which are highly relevant in the BPM landscape as many BPM suites are actually built using the Eclipse platform. The discussion relies on the seven steps of our model-driven methodology.

\begin{figure}
						\centering
						\vspace{-1em}
						\includegraphics[width=1\linewidth]{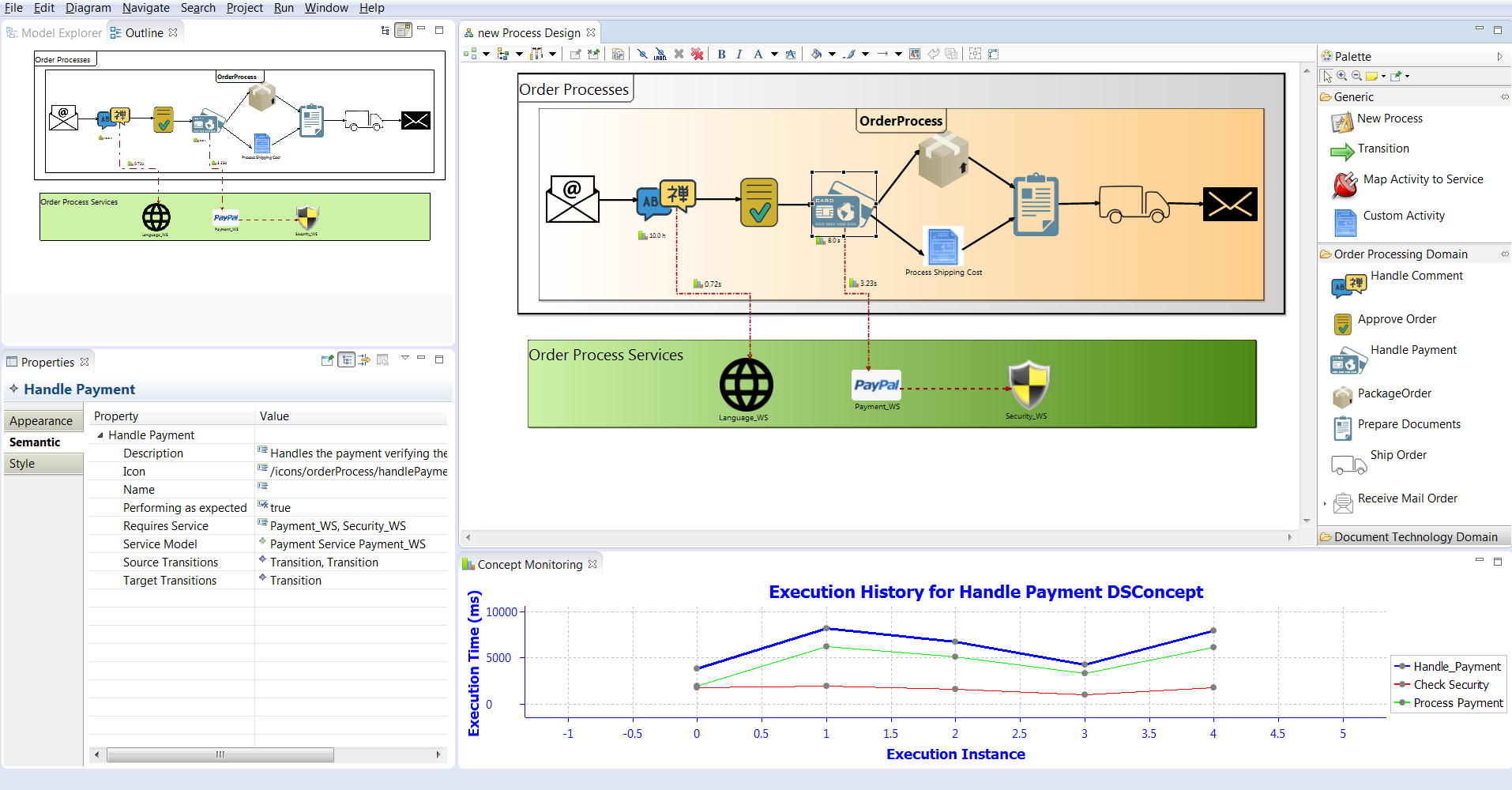}
						\vspace{-1em}
						\caption[PrototypeModeler]{Screenshot of the Eclipse-based graphical studio}
						\vspace{-1em}
						\label{fig:prototypeModeller}
					\end{figure}

	\textbf{Process design and concept mappings}. The Eclipse Modelling Framework\footnote{http://www.eclipse.org/modeling/emf/} is used for the definition of the domain-specific meta-models. Ecore meta-models are the inputs for the Sirius toolkit\footnote{https://eclipse.org/sirius/}, which allows rapid creation of graphical domain-specific modelling studios. Fig.~\ref{fig:prototypeModeller} shows a screenshot of the graphical studio. It depicts how concepts from the domain palettes can be used to compose processes that have predefined SOA connections to domain services. Monitoring information can be shown in various ways, in this particular example, execution times in the process elements indicate the BPMN activities' contribution to the overall execution time. The service contribution time is indicated in the DSConcept-Service links (dotted lines). Today, the creation of the domain-specific editor has to be supported by technical architects and developers. However, we are working on a generative approach that permits to dynamically create these modelling editors from the definition of the domain concepts.
				
	\textbf{BPMN 2.0 generation}. In addition to the Mangrove Core meta-model that we use, the Mangrove project provides a variety of plugins for model transformations as well as some editor extensions. The model-transformation plug-ins contain code that convert supported meta-models to Mangrove Core and vice-versa. They are invoked from editor plug-ins that are connected to the supported editors through standard extension points. Note that the generator only outputs the model definition and not the visual layout of the model. The BPMN~2 Modeller\footnote{https://www.eclipse.org/bpmn2-modeler/} is used to initialise the graphical representation from the generated model with a built-in Mangrove support wizard.
	
	\textbf{Enrichment of BPMN models, deployment and execution}. In our scenario, the generated BPMN skeleton is further enriched with a simple data-model, the implementation of the gateway conditions and a resource-model in order to enable execution. As Fig.~\ref{fig:prototypeArchitecture} indicates, we use the Eclipse Stardust\footnote{https://www.eclipse.org/stardust/} BPMS to execute our process. The choice of this BPMS was made because of the maturity of the tool and openness of its process monitoring API, which easily allows access to detailed process monitoring information from external components (our concept probes).

	\textbf{Monitoring}. The generation of the concept probes is done through  template instantiation. Once they are generated they need to be managed as components managed by the monitoring framework. We use the Java Management Extensions (JMX)\footnote{https://docs.oracle.com/javase/tutorial/jmx/} for our distributed monitoring infrastructure managing the probes as well as for integrating with existing monitoring frameworks. JMX is supported by a large variety of infrastructures, both commercial and open-source.
	
	As an initial validation step, we rely on the SEQUAL (SEmiotic QUALity Framework)~\cite{krogstie2006process}, which is widely used and goes beyond the modelling language to characterise its quality. We conclude that the proposed approach can significantly complement other BPMN approaches regarding the SEQUAL framework: the \emph{domain}, \emph{comprehensibility}, and \emph{organisational} appropriateness are improved by the fact that the actual focus is specific to the domain. Indeed, the framework advocates that a language must be powerful enough to express anything in the domain but no more. Also, a language should be easily extensible in order to adapt to changing business needs. These points clearly justify the interest of a DSPML on top of a BPMN model. The \emph{modeller} appropriateness and the \emph{participant} appropriateness will not change significantly as we propose to ultimately rely on BPMN. In fact, the framework recommends the use of well-known modelling languages and our approach targets basic BPMN generation. Finally, the use of proven model-driven technologies such as Sirius permit a good \emph{tool} appropriateness. Obviously, these improvements will highly depend on the proposed DSPML, but the approach provides the means to achieve them. Qualitative evaluations with final users are envisaged in order to complete the validation. Practical experiments may result in changes or refinements of the approach. The method could be extended to incorporate an user-centred approach to build the DSPML as discussed in~\cite{rieu2015participative}.

	\section{Related Work}\label{sec:relatedWork}
	
	Related work can be analysed from two main aspects: the model driven approach and the monitoring capability. Related to the \textbf{model-driven} part, Heitkotter~\cite{heitkotter2012framework} proposes DSLs4BPM, an approach for creating domain-specific process modelling languages. On the same line, Grundy et al.\cite{grundy2006generating} rely on Eclipse tooling to propose domain specific visual language editors. The difference with our approach is that these works do not provide a structured methodology to design and analyse the processes as we do. More important, the monitoring part, which is essential for business experts, is not considered. Becker et al.~\cite{becker2007picture} propose the modelling method called PICTURE, which specially focuses on public administration processes. The so-called ``process building blocks" could be compared with our domain concepts, as they are high-level domain-specific artefacts that help build the actual process. Kumaran et al.~\cite{kumaran2007using} follow the same line, proposing to automate complex and variable workflows in a service delivery management architecture. The main difference is that our approach can leverage BPMN solutions (the de-facto standard) in order to reach execution and monitoring. Other works propose extensions to BPMN 2.0 in order to be domain-specific~\cite{saeedi2010extending}. These approaches are limited by their focus on a very concrete problem space while still having to deal with the aforementioned complexity and generality of BPMN 2.0. Goal-oriented approaches~\cite{santos2010goal, lapouchnian2007requirements} use goal models as a preliminary step for process modelling. However, the graphical notations of the more popular goal oriented languages (i*, KAOS and MAP) still lack of \emph{Semantic Transparency}~\cite{moody2009physics}. There is also limited tool support for goal modelling. In addition, the goal-driven generation approaches tend to propose goal models closely tied to the business process.

	Considering the \textbf{monitoring} part, there are approaches that recur to aggregation mainly to compose events from a low-level monitoring source (using Complex Event Processing queries) in order to extract more meaningful data out of raw events~\cite{pedrinaci2008sentinel, hummer2012deriving}. Such approaches use a variety of techniques to derive better understanding of raw events, but they fundamentally still stay at a generic level with regard to the business domain. There are also approaches that try to correlate execution events to the originating processes using some forms of traceability between model elements and execution events. For instance, in~\cite{ammon2007domain}, the authors argue for the existence of domain-specific patterns for interpreting events, without giving a complete solution. Their suggestion is in line with our proposition in the idea of presenting information corresponding to domain elements, but they mostly focus on interpreting CEP events, while our approach targets structured probes that connect directly with monitoring APIs. In summary, the studied approaches recur to generic event analysis and do not provide a ``native" monitoring probe layer that directly correspond to the business concepts. To the best of our knowledge, there is no work providing an end-to-end solution for domain specific process design and monitoring.

	
	\section{Conclusion and Future Work}\label{sec:conclusion}
	
	Existing design and monitoring approaches are typically technology-specific and generic with respect to the business domain. This limits the ability of business matter experts to express their intent and enact process change. This paper leverages current BPM and SOA solutions adding a layer that is domain-dependent and platform-independent in order to facilitate process design by business matter experts. The approach presented in this paper also simplifies the management of complex business processes that span multiple domains of expertise through the support of several domain definitions during process design.
	
	We have presented a methodological and iterative approach that relies on seven main steps : 1) \emph{domain specific design} using the so-called domain concepts, which comprise the explicit representations of enterprise domain know-how; 2) \emph{the establishment of concept mappings} between domain concepts and process activities and technical services; 3) \emph{BPMN generation} relying on a pivot meta-model that enables flexibility and facilitates model transformations; 4) \emph{process enrichment}, which does not seek perfect vertical alignment between high-level models and executable ones but keeps artefacts in sync relying on concept-mappings; 5) \emph{deployment and execution}, which defines a two-step binding mechanism between domain-concepts and technical services; 6) \emph{domain specific monitoring}, based on Concept Probes and Business Process Probes that map service and process monitoring metrics  to the domain concepts and 7) the \emph{analysis} of the monitored data, which may imply the enrichment of the domain concept repository. The presented methodology is supported by tools that automate the generation and synchronisation activities. We used a mature set of open-source tools from the Eclipse Ecosystem to implement a fully functional prototype and used a running example throughout the paper to illustrate the interest and applicability of our proposition.
	
	We are focusing our next explorations on the following three main points. Firstly, the automatic generation of graphical process model editors from domain specifications mapped to the definition of the abstract syntax of the language and additional functional templates. Secondly, the integration of collaborative modelling in the aforementioned editors, which is critical in business process design. Thirdly, the automatic generation of various artefacts for the process data-model that could be used in the actual process implementation. We also aim to connect the data-model to the monitoring probes in order to correlate execution information to process data flow. These points are all under advanced stages of exploration, with a prototype being developed using Eclipse-based open-source technologies.

	\bibliography{refs}{}
	\bibliographystyle{splncs03}

\end{document}